\documentclass[journal]{IEEEtran}

\usepackage[T1]{fontenc}
\usepackage{graphicx}
\usepackage{amsmath,amsfonts,amssymb,amsthm,amstext,amscd,amsxtra,amsopn}
\usepackage{cite}
\usepackage[]{subfigure}
\usepackage{bm}
\usepackage{multirow}
\usepackage{esint}
\usepackage{empheq}
\usepackage{bm}
\usepackage{algorithm}
\usepackage{algpseudocode}
\usepackage{leftidx}
\usepackage{makecell}
\usepackage{stackrel}

\newcommand{\citeasnoun}[1]{Ref.~\citen{#1}}
\renewcommand{\Re}{\mathrm{Re}\,}
\renewcommand{\Im}{\mathrm{Im}\,}
\renewcommand{\vec}[1]{\bm{#1}}

\newcommand{\bc}[1]{\boldsymbol{\mathcal{#1}}}

\newcommand{\univec}[1]{\hat{\vec{#1}}}
\newcommand{\mat}[1]{\mathbf{#1}}
\newcommand{\matvec}[1]{\mathbf{#1}}

\def\ro{\vec{r}}
\def\rs{\vec{r}'}
\def\J{\vec{j}}
\def\M{\vec{m}}
\def\E{\vec{e}}
\def\H{\vec{h}}

\def\Ei{\vec{e}_{\rm inc}}
\def\Hi{\vec{h}_{\rm inc}}
\def\Es{\vec{e}_{\rm sca}}
\def\Hs{\vec{h}_{\rm sca}}

\def\Dp{\vec{\sigma}}

\def\ce{{c_e}}
\def\cm{{c_m}}

\newcommand{\op}[1]{\cal #1}
\newcommand{\opd}[1]{\bc{\cal{#1}} }

\newcommand\Aop{{\cal{A}}}

\newcommand\Iop{{\cal{I}}}
\newcommand\Lop{{\cal{L}}}
\newcommand\Mop{{\cal{M}}}
\newcommand\Nop{{\cal{N}}}

\newcommand\Kop{{\cal{K}}}

\newcommand\chiE{{\chi_e}}
\newcommand\chiH{{\chi_m}}

\newcommand{\intV}{\int\limits_{\rm \Omega }}
\newcommand{\intS}{\oiint\limits_{\partial \rm \Omega}}

\newcommand{\SV}[1]{{\cal V}\left({#1}\right)}

\newcommand{\ipV}[1]{\langle #1 \rangle_{{\text{\tiny V} }}}

\title{On the Computation of Power in Volume Integral Equation Formulations}

\author{Athanasios G. Polimeridis,~\IEEEmembership{Member,~IEEE}, M. T. Homer Reid, Steven G. Johnson, Jacob K. White,~\IEEEmembership{Fellow,~IEEE}, and Alejandro W. Rodriguez

\thanks{Athanasios G. Polimeridis and Jacob K. White are with the Department of Electrical Engineering and Computer Science, Massachusetts Institute of Technology.}
\thanks{M. T. Homer Reid and Steven G. Johnson are with the Department of Mathematics, Massachusetts Institute of Technology.}
\thanks{Alejandro W. Rodriguez is with the Department of Electrical Engineering, Princeton University.}}

\begin{document}

\maketitle

\begin{abstract}
We present simple and stable formulas for computing power (including absorbed/radiated, scattered and extinction power) in current-based volume integral equation formulations. The proposed formulas are given in terms of vector-matrix-vector products of quantities found solely in the associated linear system. In addition to their efficiency, the derived expressions can guarantee the positivity of the computed power. We also discuss the application of Poynting's theorem for the case of sources immersed in dissipative materials. The formulas are validated against results obtained both with analytical and numerical methods for scattering and radiation benchmark cases.
\end{abstract}

\begin{keywords}
Electromagnetic scattering, method of moments (MoM), numerical analysis, Poynting's theorem, volume integral equations.
\end{keywords}

\section{Introduction}

Volume integral equation (VIE) formulations have been extensively used over the last decades for the numerical solution of electromagnetic (EM) scattering and radiation problems (here is a non-exhaustive list of references \cite{Schaubert1984,Borup1984,Shen1989,Catedra1989,Zwamborn1992,Gan1995,Carvalho1999,Lu2003,Li2006,Rubinacci2006,Botha2006,Sancer2006,Ozdemir2006,Ozdemir2007,Sun2009,Markkanen2012}).
Admittedly, VIE methods do not hold the workhorse status of their early days, since partial differential equation (PDE)-based methods, such as the finite-difference and finite-element methods, reached a certain level of maturity. Nevertheless, some recent insights have spawned renewed interest in the development of more competitive VIE methods \cite{vanBeurden2007,vanBeurden2008,Markkanen2012b,Polimeridis2013b,Polimeridis2013d,Polimeridis2014,Peterson2014}.

The objective of this paper is to introduce concise and computationally efficient formulas for the \textit{power} absorbed, scattered, and radiated by bodies modeled with VIE methods. Of course, in the VIE or any other scattering formalism these quantities could be computed simply by direct numerical cubature of the Poynting vector over appropriate bounding surfaces, an approach we term the ``Poynting method'' (PM); however, in many cases the PM may not be the best way to perform numerical power computations. One reason is that, in methods such as the current-based VIE (JM-VIE) which solve directly for (volumetric) \textit{sources} rather than fields, computation of the Poynting vector requires an extra post-processing step to compute the fields at each cubature point. A more urgent problem is that, in many cases, the Poynting-vector cubature over the bounding surface may be badly behaved due to large cancellations from different surface regions, requiring large numbers of cubature points to obtain decent accuracy. This difficulty may be mitigated by using a distant bounding surface far removed from the surface of the scatterers, but this strategy is unavailable in many cases of interest, such as geometries involving interleaved bodies or problems in which we need the power absorbed by just one of two nearby objects.

These issues were recently discussed in the context of \textit{surface} integral equation (SIE) solvers \cite{Kern2010,Reid2014}. More specifically, \citeasnoun{Kern2010} noted that the power absorbed by a homogeneous body may be computed solely from knowledge of the tangential currents flowing on its surface. \citeasnoun{Reid2014} extended this observation by noting that in fact the surface currents suffice to determine not only the absorbed power but also the scattered power, total power (extinction), force (radiation pressure), and torque on a homogeneous body. Indeed, as discussed in \cite{Reid2014}, all of these quantities may be expressed compactly as vector-matrix-vector product (VMVP) expressions of the form $Q=\mathbf{c}^\ast \mathbf{M} \mathbf{c}$, where $\mathbf{c}$ is the vector of surface-current basis-function coefficients obtained in the SIE solution to a scattering problem and $\mathbf{M}$ is a matrix which assumes different forms for the various different quantities $Q$ we are calculating. In this work we extend the developments of \cite{Reid2014} to the VIE domain, deriving VMVP formulas for the absorbed, scattered, radiated, and total power. Our formulas are analogous to those of \cite{Reid2014} in that they compute the power directly from the volume-current coefficient vector (the solution to the linear VIE system), thus bypassing the post-processing step of computing scattered fields and Poynting's vectors. An advantage of our formulas over their SIE counterparts is that, as we prove rigorously in Section IIIE, the numerical predictions that they yield obey the key physical requirement of positivity of the power absorbed, scattered\footnote{As explained in Section IIIC, an additional matrix vector product may be needed.}, and radiated by passive material bodies, even in their discretized form. In contrast, while the SIE power formulas derived in \cite{Reid2014} (see also \cite{Rodriguez2013}) are positive in the exact Maxwell equations, this positivity relied on a delicate cancellation that can sometimes break down when they are approximated by a discretized basis (such as boundary elements).

As noted above, the VMVP formulas derived previously in the SIE context for \textit{classical} scattering \cite{Reid2014} have also proven useful for numerical modeling of \textit{quantum/statistical-mechanical}
phenomena \cite{Rodriguez2013,Reid2013,scuff1}. In direct analogy, the formulas presented here for classical problems are key building blocks for an efficient VIE-based numerical approach to computational fluctuation physics; in particular, our VMVP formulas for the absorbed power \eqref{P_abs} and the radiated power \eqref{P_rad_final2} may be extended to matrix-trace formulas for rates of radiative heat transfer and fluorescence. This correspondence will be addressed in future work.

The remainder of this paper is organized as follows. In Section II we set up the JM-VIE formulation and the associated linear system. In addition, we describe the incorporation of dipole sources in JM-VIE solvers. In Section III, we present the main results of this work: the boxed VMVP expressions for the computation of power in VIE methods, and we prove their positivity. Finally, in Section IV we validate our formulas, and we demonstrate some of their useful properties. Table I lists some notation used in this work.

\begin{table}[ht]
\caption{Notation} \label{table_notation} \centering
\begin{tabular}{c| l }
\hline\hline\\[-0.4em]
Notation                               & Description                                                              \\[0.4em] \hline \\[-0.4em]
$\vec{a}$                              & vector in $\mathbb{C}^3$, $\vec{a} = |\vec{a}\! |\, \univec{a} =(a_x,a_y,a_z)$              \\[0.3em]
$\matvec{a}$                          & one-dimensional array (vector in $\mathbb{C}^N$)              \\[0.3em]
$\mat{A}$                             & matrix in $\mathbb{C}^{n_1\times n_2}$     \\[0.3em]
$\overline{\vec{a}}$                  & complex conjugate           \\[0.3em]
$\matvec{a}^\ast$                     & conjugate transpose             \\[0.3em]
$\op{L}$ $(\opd{L}) $                      & operator acting on vectors in $\mathbb{C}^3$ ($\mathbb{C}^6$) \\[0.3em]
\hline\hline
\end{tabular}
\end{table}

\section{Volume integral equations}

\subsection{Formulation}

We consider the scattering of time-harmonic EM waves by a penetrable object, occupying the bounded domain $\rm \Omega$ in 3-D Euclidean space, $\mathbb{R}^3 $. The working angular frequency is $\omega\in \mathbb{R}^+$ and the electric properties are defined as
\begin{equation}
\begin{split}
\epsilon &= \epsilon_0,\, \mu=\mu_0 \quad \text{in}\,\, \mathbb{R}^3 \backslash \Omega ; \\
\epsilon &= \epsilon_r(\ro)\,\epsilon_0,\, \mu=\mu_r(\ro)\mu_0 \quad \text{in}\,\,  \Omega
\end{split}
\end{equation}
Here, the vacuum (or free-space) permittivity $\epsilon_0$ and permeability $\mu_0$ are real positive values, while the relative permittivities $\epsilon_r(\ro)$ and $\mu_r(\ro)$ read
\begin{equation}
\begin{split}
\epsilon_r(\ro) &= \epsilon_r'(\ro) - i \epsilon_r''(\ro)\\
\mu_r(\ro)      &= \mu_r'(\ro) - i \mu_r''(\ro)
\end{split}
\end{equation}
with $i=\sqrt{-1}$ and $\epsilon_r'', \mu_r''\in [0,\infty) $, assuming a time factor $\exp{(i\omega t)}$.

The total time harmonic fields ($\E,\H$) in the presence of an isotropic inhomogeneous object can be expressed in terms of equivalent polarization and magnetization currents ($\J,\M$), as follows (dropping some function arguments where no confusion exists):
\begin{equation}\label{ScatteringTheorem}
\begin{pmatrix} \E \\ \H \end{pmatrix}
=
\begin{pmatrix} \Ei \\ \Hi \end{pmatrix}
+
\begin{pmatrix} \Es \\ \Hs \end{pmatrix}
\end{equation}
where the incident fields ($\Ei,\Hi$) are the fields generated by sources in the absence of the scatterer and the scattered fields ($\Es,\Hs$) are given by\footnote{More on the use of $\Nop$ operator can be found in \cite{Sun2009,Markkanen2012b,Polimeridis2014}.}
\begin{equation}\label{A_sca}
\begin{split}
\begin{pmatrix} \Es \\ \Hs \end{pmatrix}
&=
\begin{pmatrix}
 \frac{1}{\ce} \Lop & - \Kop \\
  \Kop &  \frac{1}{\cm} \Lop
\end{pmatrix}
\begin{pmatrix} \J \\ \M \end{pmatrix}\\[0.5em]
&=
\underbrace{\begin{pmatrix}
 \frac{1}{\ce} (\Nop-\Iop) & - \Kop \\
  \Kop &  \frac{1}{\cm} (\Nop-\Iop)
\end{pmatrix}}_{\opd{A}_{\rm sca}}
\begin{pmatrix} \J \\ \M \end{pmatrix}
\end{split}
\end{equation}
where $\ce,\cm \triangleq i\omega\epsilon_0, i\omega\mu_0$. $\opd{A}_{\rm sca}$ is simply the convolution operator with the $6\times 6$ Green function connecting currents to fields in vacuum.   More explicitly, the associated integro-differential operators are
\begin{subequations}
\begin{align}
\Lop \vec{f} &\triangleq (k_0^2 + \nabla\nabla\cdot) \SV{ \vec{f}}\\
\Kop \vec{f} &\triangleq \nabla\times \SV{ \vec{f}}\\
\Nop \vec{f} &\triangleq \nabla\times\nabla\times\SV{\matvec{f}}
\end{align}
\end{subequations}
where
\begin{equation}
\SV{\vec{f}} \triangleq \intV g(\ro-\rs) \vec{f}(\rs)d^3\rs
\end{equation}
is the volume vector potential and $g$ is the fundamental Helmholtz solution,
\begin{equation}
g(\ro) = \frac{e^{-i k_0 |\ro|}}{4 \pi |\ro|}
\end{equation}
with $k_0=\omega\sqrt{\epsilon_0\mu_0}$ being the wavenumber in free-space. Also, the equivalent current densities are defined in terms of the fields as follows:
\begin{subequations}\label{JM}
\begin{align}
\J(\ro)  &\triangleq \ce \chiE(\ro)\,  \E(\ro) \\
\M(\ro)  &\triangleq \cm     \chiH(\ro)\,  \H(\ro).
\end{align}
\end{subequations}
where
\begin{equation}
\chiE\triangleq\epsilon_r-1,\quad \chiH\triangleq\mu_r-1
\end{equation}
is the electric and magnetic susceptibility, respectively. Finally, the JM-VIE formulation can be derived by combining \eqref{ScatteringTheorem}, \eqref{A_sca} and \eqref{JM}  \cite{Markkanen2012b,Polimeridis2014},
\begin{equation}\label{JMVIE}
\opd{A}
\begin{pmatrix}
\J     \\
\M
\end{pmatrix}
=
\opd{C}\opd{M}_{\chi}
\begin{pmatrix}
\Ei     \\
\Hi
\end{pmatrix}
\end{equation}
where
\begin{equation}
\begin{split}
\opd{A} &=
\begin{pmatrix}
 \Aop_e^{\rm N}   & \Aop_e^{\rm K} \\[0.5em]
- \Aop_m^{\rm K}                 &  \Aop_m^{\rm N}
\end{pmatrix}\\[0.5em]
&= \begin{pmatrix}
 \Mop_{\epsilon_r} - \Mop_{\chi_e} \Nop   & c_e \Mop_{\chi_e} \Kop \\[0.5em]
- c_m \Mop_{\chi_m} \Kop                 &  \Mop_{\mu_r} - \Mop_{\chi_m} \Nop
\end{pmatrix}
\end{split}
\end{equation}
and
\begin{equation}
\opd{M}_{\chi}
=
\begin{pmatrix}
\Mop_{\chi_e} & 0    \\
0 & \Mop_{\chi_m}
\end{pmatrix},\quad
\opd{C}
=
\begin{pmatrix}
\ce \op{I} & 0    \\
0 & \cm \op{I}
\end{pmatrix}.
\end{equation}
$\Mop_{\phi}$ are multiplication operators that multiply by the local parameter functions $\phi$, while $\op{I}$ is the identity dyadic tensor.

\subsection{Linear System}

Usually, JM-VIE formulations are numerically solved by means of a Galerkin method, where the equivalent volumetric currents are approximated as expansions in some discrete set of vector-valued square-integrable basis functions, e.g. $\vec{p} \in [L^2(\mathbb{R}^3)]^3$ as in \cite{vanBeurden2008,Markkanen2012b,Polimeridis2014}:
\begin{equation}\label{JM_exp}
\J \approx \sum\limits_{\alpha} x_{e,\alpha} \vec{p}_{\alpha},\quad
\M \approx \sum\limits_{\alpha} x_{m,\alpha} \vec{p}_{\alpha}
\end{equation}
The linear system arising from the Galerkin ``testing'', i.e.  $\mat{A}\matvec{x}=\matvec{b}$ with $\matvec{x},\matvec{b}\in\mathbb{C}^{N}$ and $\mat{A}\in\mathbb{C}^{N\times N}$, reads
\begin{equation}\label{system_matrix:eq}
\begin{pmatrix}
 \mat{A}_e^{\rm N}   & \mat{A}_e^{\rm K} \\[0.5em]
- \mat{A}_m^{\rm K}                 &  \mat{A}_m^{\rm N}
\end{pmatrix}
\begin{pmatrix}
\matvec{x}_e     \\
\matvec{x}_m
\end{pmatrix}
=
\begin{pmatrix}
\matvec{b}_e     \\
\matvec{b}_m
\end{pmatrix}
\end{equation}
where
\begin{equation}
\begin{split}
\mat{A}&=
\begin{pmatrix}
 \mat{A}_e^{\rm N}   & \mat{A}_e^{\rm K} \\[0.5em]
- \mat{A}_m^{\rm K}                 &  \mat{A}_m^{\rm N}
\end{pmatrix}\\[0.5em]
&=
\begin{pmatrix}
\mat{M}_{\epsilon_r} \mat{G} - \mat{M}_{\chi_e} \mat{N}   & c_{e} \mat{M}_{\chi_e} \mat{K} \\[0.5em]
-c_{m} \mat{M}_{\chi_m} \mat{K}                 &  \mat{M}_{\mu_r} \mat{G} - \mat{M}_{\chi_m} \mat{N}
\end{pmatrix}
\end{split}
\end{equation}
with
\begin{subequations}\label{KN_op}
\begin{align}
\mat{N}_{\alpha\beta} &= \ipV{\vec{p}_{\alpha}, \Nop \vec{p}_{\beta}} \\
\mat{K}_{\alpha\beta} &= \ipV{\vec{p}_{\alpha}, \Kop \vec{p}_{\beta}}
\end{align}
\end{subequations}
and $\mat{G}$ is the Gram matrix, given by
\begin{equation}
\mat{G}_{\alpha\beta} =  \ipV{\vec{p}_{\alpha},  \vec{p}_{\beta}}.
\end{equation}
Also, $\mat{M}$ and $\mat{C}$ are the discrete versions of the associated operators. More specifically, matrices $\mat{M}$ are diagonal for isotropic material with the non-zero values being equal to the material properties at the corresponding element, while matrix $\mat{C}$ is the identity matrix with multiplication pre-factors $\ce$ and $\cm$ for the two diagonal sub-blocks, respectively. Note that the equivalent currents have no continuity constraints (at the interface of the elements) and the support of the basis/testing functions is restricted to single elements. Hence, the associated Gram matrix is diagonal, when non-overlapping basis functions are used \cite{Markkanen2012b,Polimeridis2014}.

In the above we have used the inner product:
\begin{equation}
\ipV{\vec{f},  \vec{g}} = \intV{\overline{\vec{f}} \cdot  \vec{g}}.
\end{equation}
Finally, the right-hand side in \eqref{system_matrix:eq} is given by
\begin{align}\label{rhs}
\matvec{b} = \begin{pmatrix} \matvec{b}_{e} \\ \matvec{b}_{m}  \end{pmatrix}
= \mat{C} \mat{M}_{\chi} \begin{pmatrix} \matvec{e}_{\rm inc} \\ \matvec{h}_{\rm inc}  \end{pmatrix}
\end{align}
where
\begin{subequations}\label{inc_test}
\begin{align}
\matvec{e}_{{\rm inc},\alpha} &= \ipV{\vec{p}_{\alpha},\vec{e}_{\rm inc}} \\
\matvec{h}_{{\rm inc},\alpha} &= \ipV{\vec{p}_{\alpha},\vec{h}_{\rm inc}}.
\end{align}
\end{subequations}

\subsection{Dipole Excitation}\label{voxeldipole}

In most radiation problems, we need to deal with elementary excitations, such as electric (or magnetic) Hertzian oscillating dipoles, $\vec{d}(\ro)=  \delta^3(\ro)\hat{\vec{d}}$. Numerically, the finite discretization means that the 3D dirac delta will have support only within a voxel of size $\Delta V$, and will be given by
\begin{equation}\label{limit}
\delta_{\triangle V}(\ro) = \frac{p}{\triangle V} =
\begin{cases}
\frac{1}{\triangle V}, \quad  \text{ if $\ro \in {\rm supp}(p)$}\\
0, \quad \text{otherwise}
\end{cases}
\end{equation}
with $p$ being the magnitude of the basis functions for the approximation of the polarization/magnetization densities.

The use of dipole sources in our analysis could prove quite problematic, since current sources are inherently modeled in VIE as uniform distributions throughout the volume elements. Motivated by the limit $\delta^3(\ro) = \lim\limits_{\triangle V \rightarrow 0+}\delta_{\triangle V}(\ro)$, we introduce the notion of the \textit{distributed-dipole} (DD) source. Specifically, the current of a DD source immersed in element $\alpha$, along the direction of the $\vec{p}_{\alpha}$ basis reads
\begin{equation}
\Dp_{\alpha}(\vec{x})=  \frac{\vec{p}_{\alpha}}{V_\alpha}.
\end{equation}
Hence, the impressed current vector is given by
\begin{equation}\label{d_vector}
\begin{pmatrix} \matvec{d}_{e} \\ \matvec{d}_{m}  \end{pmatrix} = \hat{\mat{G}}^{-1} \begin{pmatrix} \matvec{p}_{e} \\ \matvec{p}_{m}  \end{pmatrix}
\end{equation}
where
\begin{equation}
\hat{\mat{G}}=
\begin{pmatrix}
\mat{G} & 0\\
0 & \mat{G}
\end{pmatrix}.
\end{equation}
The non-zero elements of the vector in \eqref{d_vector} depend solely on the location and direction of the dipole sources under consideration.

Finally, the incident fields of \eqref{rhs} read
\begin{equation}
\begin{pmatrix} \Ei \\ \Hi \end{pmatrix} =
\mat{A}_{\rm sca}  \begin{pmatrix} \matvec{d}_{e} \\ \matvec{d}_{m}  \end{pmatrix}
\end{equation}
where $\mat{A}_{\rm sca}$ is the discrete form of the operator in \eqref{A_sca}:
\begin{equation}\label{A_sca_discrete}
\mat{A}_{\rm sca}=
\begin{pmatrix}
 \frac{1}{\ce}(\mat{N} - \mat{G})   & -\mat{K} \\[0.5em]
\mat{K}                 &   \frac{1}{\cm}(\mat{N} - \mat{G})
\end{pmatrix}.
\end{equation}
In the case of dipole sources located outside the scatterer, the computation of the incident fields is quite straightforward, since there are no singularities in the fields (inside the scatterer). More specifically, the impressed sources are propagated by means of the free-space Green function, and the incident fields are ``tested'' as in \eqref{inc_test}.

\section{Power Formulas}\label{PowerFormulas}

The power flowing into or radiated from a material body $\rm \Omega$ can be expressed as an integral of the normal (inward or outward-directed, respectively) component of the (total) Poynting vector over its surface $\partial\rm \Omega$ \cite{Jackson_book}. Similar expressions may also be derived for the scattered and extinguished power by integrating the associated components of the Poynting vector. In this section, we present  simple formulas for computing the powers directly from the equivalent polarization and magnetization currents, which are the immediate output of the numerical solution of JM-VIE formulations. The derivation is based on the conservation of energy, or Poynting's theorem, which relates the energy flowing out through the boundary surfaces of the body to the work done by the fields on the currents \cite{Jackson_book}.

In particular, we find it useful to decompose the time average Poynting vector \cite{Jackson_book}
\begin{equation}
\langle\vec{s}_{\rm tot}\rangle \triangleq \frac{1}{2}\Re{}\left( \vec{e}_{\rm tot}\times \overline{\vec{h}}_{\rm tot} \right)
\end{equation}
in terms of incident, scattered, and extinguished components:
\begin{equation}
\langle\vec{s}_{\rm tot}\rangle = \langle\vec{s}_{\rm inc}\rangle + \langle\vec{s}_{\rm sca}\rangle + \langle\vec{s}_{\rm ext}\rangle
\end{equation}
where
\begin{subequations}
\begin{align}
\langle\vec{s}_{\rm inc}\rangle &\triangleq \frac{1}{2}\Re{}\left( \Ei \times \overline{\vec{h}}_{\rm inc} \right)  \\
\langle\vec{s}_{\rm sca}\rangle &\triangleq \frac{1}{2}\Re{}\left( \Es \times \overline{\vec{h}}_{\rm sca} \right)  \\
\langle\vec{s}_{\rm ext}\rangle &\triangleq \frac{1}{2}\Re{}\left( \Es \times \overline{\vec{h}}_{\rm inc}  + \Ei \times \overline{\vec{h}}_{\rm sca} \right)
\end{align}
\end{subequations}
are given by the corresponding fields. Poynting's theorem can therefore be written in the following form\footnote{Note that $ \nabla \cdot \langle\vec{s}_{\rm inc}\rangle=0$ for a lossless ambient medium.} \cite{Jackson_book}:
\begin{equation}\label{cons_law}
\nabla \cdot\langle\vec{s}_{\phi}\rangle = - \langle W_{\phi}\rangle
\end{equation}
where
\begin{equation}
\langle W_{\phi}\rangle =  \langle W_{\phi}^{f}\rangle + \langle W_{\phi}^{b}\rangle
\end{equation}
with $\phi \in \{{\rm inc,sca,tot} \}$, is the time average of the work done on total currents by the corresponding fields. More specifically, it is the sum of work done on free currents ($W_{\phi}^{f}$) and bound (polarization and magnetization) currents ($W_{\phi}^{b}$):
\begin{equation}
\langle W_{\phi}^{f,b}\rangle \triangleq  \frac{1}{2}\Re{} \left( \, \overline{\J}_{f,b}\cdot \E_{\phi} + \overline{\M}_{f,b} \cdot \H_{\phi}  \right).
\end{equation}
In the next few subsections we derive the formulas for the computation of absorbed, extinction, scattered, and radiated power, along with a simple proof for their positivity.

\subsection{Absorbed Power}

The absorbed power is the power flowing into the body and is given by the integral of the inward-directed normal component of the Poynting vector. With the help of \eqref{cons_law} and the divergence theorem, we derive the absorbed power in terms of volumetric quantities:
\begin{equation}\label{W_tot}
\begin{split}
   P_{\rm abs} &= - \intS \langle\vec{s}_{\rm tot}\rangle\cdot \univec{n} = \intV\langle W_{\rm tot}\rangle\\
   &= \frac{1}{2}\Re{\left( \ipV{\J, \E_{\rm tot}} + \ipV{\M, \H_{\rm tot}} \right)}
\end{split}
\end{equation}
where $\univec{n}$ is the outward-directed surface normal. Now, inserting the current expansions \eqref{JM_exp} in the inner products of \eqref{W_tot}, we get
\begin{equation}\label{W_tot2}
\begin{split}
& \ipV{\J, \frac{\J}{\ce \chiE}} + \ipV{\M, \frac{\M}{\cm \chiH}}\\
 &= \frac{1}{\,\ce}\sum\limits_{\alpha \beta} x_{e,\alpha}^{\ast} \frac{\ipV{\vec{p}_{\alpha},  \vec{p}_{\beta}} }{\chi_{e,\alpha\beta}} x_{e,\beta} \\
 &+ \frac{1}{\,\cm}\sum\limits_{\alpha \beta} x_{m,\alpha}^{\ast} \frac{\ipV{\vec{p}_{\alpha},  \vec{p}_{\beta}} }{\chi_{m,\alpha\beta}} x_{m,\beta} \\
&= \frac{1}{\ce} \matvec{x}_e^{\ast} \left( \mat{M}_{\chi_e}^{-1} \mat{G} \right) \matvec{x}_e
+ \frac{1}{\cm} \matvec{x}_m^{\ast} \left( \mat{M}_{\chi_m}^{-1} \mat{G} \right) \matvec{x}_m\\
&= \matvec{x}^{\ast}\hat{\mat{M}}\hat{\mat{G}}\matvec{x}
\end{split}
\end{equation}
where
\begin{equation}\label{hatM}
\hat{\mat{M}}= (\mat{C}\, \mat{M}_{\chi})^{-1}.
\end{equation}
In \eqref{W_tot2}, we have used the definition of the equivalent polarization currents \eqref{JM} in order to replace the total fields. The substitution is admissible for nonzero susceptibilities, otherwise the currents are identically zero and there is no contribution to the total work. Finally, the absorbed power takes the form
\begin{equation}\label{P_abs}
\boxed{
P_{\rm abs} =  \frac{1}{2}\Re{} \matvec{x}^{\ast}\hat{\mat{M}}\hat{\mat{G}}\matvec{x}.
}
\end{equation}

\subsection{Extinction Power}

The extinction power is the total power removed from the incident field (the sum of the absorbed and the scattered powers) due to the presence of the scattering object $\rm \Omega$, and is given by similar considerations as in the absorbed power computation, as follows:
\begin{equation}
\begin{split}
   P_{\rm ext} &= - \intS \langle\vec{s}_{\rm ext} \rangle \cdot \univec{n}\\
   & = \intV\langle W_{\rm inc} \rangle = \frac{1}{2}\Re{\left( \ipV{\J, \Ei} + \ipV{\M, \Hi} \right)}.
\end{split}
\end{equation}
The computation of the work done by the incident fields on the polarization and magnetization currents can be simply expressed in terms of quantities from the linear system of JM-VIE solution. More specifically, the incident fields are related to the right hand side vector as shown in \eqref{rhs}, and the associated inner products admit the following representation:
\begin{equation}
\begin{split}
& \ipV{\J, \frac{\vec{b}_e}{\ce \chiE}} + \ipV{\M, \frac{\vec{b}_m}{\cm \chiH}}\\
&= \frac{1}{\ce}\sum\limits_{\alpha} x_{e,\alpha}^{\ast} \frac{\ipV{\vec{p}_{\alpha},  \vec{b}_{e}} }{\chi_{e,\alpha\beta}}\\
&+ \frac{1}{\cm}\sum\limits_{\alpha} x_{m,\alpha}^{\ast} \frac{\ipV{\vec{p}_{\alpha},  \vec{b}_{m}} }{\chi_{m,\alpha\beta}} \\
&= \frac{1}{\ce} \matvec{x}_e^{\ast}\,  \mat{M}_{\chi_e}^{-1} \matvec{b}_e
+ \frac{1}{\cm} \matvec{x}_m^{\ast}\,  \mat{M}_{\chi_m}^{-1} \matvec{b}_m\\
&= \matvec{x}^{\ast}\hat{\mat{M}}\matvec{b}.
\end{split}
\end{equation}
The final formula for the extinction power reads
\begin{equation}\label{P_ext}
\boxed{
P_{\rm ext} = \frac{1}{2}\Re{} \matvec{x}^{\ast}\hat{\mat{M}}\matvec{b}.
}
\end{equation}

It is also useful to write $P_{\rm ext}$ as the real part of an analytic/causal function (see Appendix), both from a theoretical perspective (to get an analogue of the optical theorem as in \cite{Hashemi2012}) and from a practical perspective (e.g. for transforming frequency averaging into a complex frequency, as in \cite{Liang2013,Miller2014}).

\subsection{Scattered Power}\label{P_sca_Section}

The power scattered from an object $\rm \Omega$ is given by the real part of the integral of the outward-directed normal component of $\vec{s}_{\rm sca}$ over $\partial \rm \Omega$\footnote{Note here the plus sign!}:
\begin{equation}
P_{\rm sca} = + \intS \langle\vec{s}_{\rm sca}\rangle \cdot \univec{n}.
\end{equation}
Obviously, the scattered power can be expressed in terms of quantities arising in JM-VIE linear system, as the difference between the extinction power and the absorbed power:
\begin{equation}\label{P_sca1}
\begin{aligned}
P_{\rm sca} &= P_{\rm ext} - P_{\rm abs}\\
 &= \frac{1}{2}\Re{} \matvec{x}^{\ast}\hat{\mat{M}}\left(  \matvec{b} - \hat{\mat{G}}\matvec{x} \right).
\end{aligned}
\end{equation}
In cases where the scattering mechanism is weak compared to absorption, formula \eqref{P_sca1} may be prone to numerical instabilities, i.e., computing a small number as the difference of two almost equal and possibly large approximate values. Therefore, it would be useful to derive some additional formulas for this case, that are immune to numerical instabilities. In doing so, we resort again to the conservation laws \eqref{cons_law}:
\begin{equation}
\begin{split}
 \intS \langle\vec{s}_{\rm sca} \rangle \cdot \univec{n} = -\intV\langle W_{\rm sca} \rangle
\end{split}
\end{equation}
where
\begin{equation}\label{Wsca}
W_{\rm sca}=\ipV{\J, \Es} + \ipV{\M, \Hs}.
\end{equation}
The first term of the right hand side is given by
\begin{equation}\label{Wjsca}
\begin{split}
&\ipV{\J, \Es} =  \ipV{\J, \frac{1}{\ce } (\Nop\J-\J)} - \ipV{\J, \Kop\M} \\
&= \frac{1}{\,\ce}\sum\limits_{\alpha \beta} x_{e,\alpha}^{\ast} \left( \ipV{\vec{p}_{\alpha}, \Nop \vec{p}_{\beta}} - \ipV{\vec{p}_{\alpha},  \vec{p}_{\beta}} \right) x_{e,\beta} \\
&- \sum\limits_{\alpha \beta} x_{e,\alpha}^{\ast}  \ipV{\vec{p}_{\alpha}, \Kop \vec{p}_{\beta}} x_{m,\beta}\\
&= \frac{1}{\ce}   \matvec{x}_e^{\ast}\,  \left( \mat{N} - \mat{G} \right) \matvec{x}_e
-\matvec{x}_e^{\ast}\,   \mat{K} \matvec{x}_m
\end{split}
\end{equation}
and with similar considerations, the second term reads
\begin{equation}\label{Wmsca}
\begin{split}
\ipV{\M, \Hs} =  \frac{1}{\cm}   \matvec{x}_m^{\ast}\,  \left( \mat{N} - \mat{G} \right) \matvec{x}_m
+\matvec{x}_m^{\ast}\,   \mat{K} \matvec{x}_e.
\end{split}
\end{equation}
Hence, combining \eqref{Wsca}, \eqref{Wjsca} and \eqref{Wmsca} with \eqref{A_sca_discrete}, we get
\begin{equation}
W_{\rm sca}=
\matvec{x}^{\ast}\mat{A}_{\rm sca}\matvec{x}.
\end{equation}
Finally, we obtain an alternative formula for the computation of the scattered power:
\begin{equation}\label{P_sca2}
\boxed{
P_{\rm sca} =  -\frac{1}{2}\Re{} \matvec{x}^{\ast}\mat{A}_{\rm sca}\matvec{x}.
}
\end{equation}
As mentioned above, there is a trade-off in numerical complexity for getting more stable formula: the matrix in \eqref{P_sca2} is dense, so the cost of the numerical evaluation scales like ${\cal{O}}(N^2)$ instead of the ${\cal{O}}(N)$ scaling of \eqref{P_sca1}. Note, though, that JM-VIE formulations typically result in very large linear systems and fast solvers are employed for their numerical solution. In this case, the complexity of evaluating \eqref{P_sca2} scales like ${\cal{O}}(N\log{N})$ (\cite{Rokhlin1990,Phillips1997,Lu2003,Polimeridis2014} among others), and the associated operators in \eqref{A_sca_discrete} have been pre-computed in the actual numerical solution of the JM-VIE linear system.

\subsection{Radiated Power}

Here we consider radiation from sources immersed in $\rm \Omega$, in particular from elementary sources, i.e., electric and magnetic point Hertzian dipoles. The power radiated from $\rm \Omega$ is given by
\begin{equation}\label{P_rad1}
   P_{\rm rad} =  \intS \langle\vec{s}_{\rm tot}\rangle\cdot \univec{n}.
\end{equation}
Obviously, the total fields generated in this case are singular at the location of the sources, hence we resort to the natural generalization of the divergence theorem, where the derivatives are taken in the weak/distribution sense,
\begin{equation}\label{P_rad2}
\begin{split}
   P_{\rm rad} =  \intS \langle\vec{s}_{\rm tot}\rangle\cdot \univec{n}
   &= - \intV\langle W^{f}_{\rm tot} \rangle - \intV\langle W^{b}_{\rm tot} \rangle\\
   &= P_{\rm sup} - P_{\rm abs}
\end{split}
\end{equation}
where
\begin{equation}
\begin{split}
   P_{\rm sup} &= - \intV\langle W^{f}_{\rm tot} \rangle\\
   &= -  \frac{1}{2}\Re{\left( \ipV{\J_f, \E_{\rm tot}} + \ipV{\M_f, \H_{\rm tot}} \right)}
\end{split}
\end{equation}
is the power supplied by the source and $P_{\rm abs}$ the absorbed power in $\Omega$, already defined. Note, that in the case of dissipative media, both supplied and absorbed powers are infinite \cite{Tai2000} (a problem related also to the ill-defined local density of states \cite{Scheel1999,VanVlack2012}). Nevertheless, the radiated power (i.e., power flowing from the surface of $\Omega$) is still a finite quantity and represents the outward power flow from a dipole source with constant amplitude. Otherwise, the notion of the ``insulated'' dipole could be used (as in \cite{Tai2000}), especially when the actual supplied power or the efficiency of the radiator is under scrutiny.

The supplied power can be easily derived with the help of the DD source,
\begin{equation}\label{Psup}
\begin{split}
P_{\rm sup} &= P_{{\rm sup}_{\rm inc}} + P_{\rm sup_{\rm sca}} \\
&=-\frac{1}{2}\Re{} \left\{ \matvec{d}^{\ast} \mat{A}_{\rm sca} \matvec{d}
+ \matvec{d}^{\ast}\mat{A}_{\rm sca} \matvec{x} \right\}
\end{split}
\end{equation}
and the radiated power formula reads
\begin{equation}\label{P_rad_final}
P_{\rm rad} = \frac{1}{2}\Re{} \left\{  -\matvec{d}^{\ast} \mat{A}_{\rm sca} ( \matvec{d}+ \matvec{x})
- \matvec{x}^{\ast}\hat{\mat{M}}\hat{\mat{G}}\matvec{x}   \right\}.
\end{equation}
As with \eqref{P_sca1}, the above formula is prone to catastrophic cancellations, especially considering that the values there could be very large. An alternative (and more intuitive) formula for the radiated power can be obtained by expanding the fields also for the case of the absorbed power,
\begin{equation}\label{Pabs_DD}
\begin{split}
P_{\rm abs} &= P_{\rm ext} - P_{\rm sca} \\
&=\frac{1}{2}\Re{} \left\{  \matvec{x}^{\ast} \hat{\mat{M}}\matvec{b}
+ \matvec{x}^{\ast}\mat{A}_{\rm sca}\matvec{x}   \right\}\\
&=\frac{1}{2}\Re{} \left\{  \matvec{x}^{\ast} \hat{\mat{M}} \hat{\mat{M}}^{-1} \mat{A}_{\rm sca}\matvec{d}
+ \matvec{x}^{\ast}\mat{A}_{\rm sca}\matvec{x}   \right\}
\\
&=\frac{1}{2}\Re{}   \matvec{x}^{\ast} \mat{A}_{\rm sca} ( \matvec{d} + \matvec{x}).
\end{split}
\end{equation}
where $\matvec{d}$ is defined in \eqref{d_vector}. Hence, by combining \eqref{Psup} and \eqref{Pabs_DD} the radiated power admits an elegant quadratic form:
\begin{equation}\label{P_rad_final2}
\boxed{
P_{\rm rad} = -\frac{1}{2}\Re{} (\matvec{x}+\matvec{d})^{\ast} \mat{A}_{\rm sca} ( \matvec{x}+ \matvec{d}).
}
\end{equation}
Interestingly, the derived quadratic power formula has the same computation complexity as \eqref{P_rad_final}.

\subsection{Positivity}

In addition to their efficiency and simplicity, the above formulas manifest the positivity of power in passive media in a numerically stable fashion: positivity is preserved by discretization. As observed in \cite{Kern2010}, this is not always the case for SIE formulas. Similar behavior is also expected from the difference formulas \eqref{P_sca1} and \eqref{P_rad_final} when the parts have almost equal values. While those formulas are analytically exact, their potential reliance on a large cancellation to leave a positive remainder makes them susceptible to numerical inaccuracy and a loss of positivity when they are approximated via a discretized basis. In contrast, we show here that our scattered \eqref{P_sca2} and radiated \eqref{P_rad_final} power formulas are expressed in terms of manifestly positive-definite quadratic forms, and hence this positivity is preserved by any Galerkin discretization.

We begin with the notion of the Hermitian decomposition: Every matrix $\mat{B}$ can be decomposed into the Hermitian ($\mat{B}^{\rm H}$) and the skew-Hermitian ($\mat{B}^{\rm SH}$) components, as follows:
\begin{equation}
\mat{B} = \mat{B}^{\rm H} + \mat{B}^{\rm SH}
\end{equation}
where
\begin{equation}
\begin{aligned}
 \mat{B}^{\rm H}   &= (\mat{B}^{\rm H})^{\ast} =    \frac{\mat{B}+\mat{B}^{\ast}}{2} \\[0.5em]
  \mat{B}^{\rm SH} &= -(\mat{B}^{\rm SH})^{\ast} = \frac{\mat{B}-\mat{B}^{\ast}}{2}.
 \end{aligned}
\end{equation}
Obviously, the quadratic forms of these components are purely real and imaginary, respectively.
Since, in all formulas we are computing the real part of the associated quadratic forms, the positivity is guaranteed if the Hermitian component is positive- or negative-semidefinite, depending on the sign of the final formula. Hence, the sufficient conditions read
\begin{equation}
\begin{split}
\Re{} \matvec{x}^{\ast}\mat{B}\matvec{x} &= \matvec{x}^{\ast}\mat{B}^{\rm H}\matvec{x} \geq 0\\
 \text{iff}\,\, &\mat{B}^{\rm H} \succeq 0.
\end{split}
\end{equation}

In the case of the absorbed power formula \eqref{P_abs},
\begin{equation}
\mat{B}^{\rm H} = \hat{\mat{G}} \hat{\mat{M}}^{\rm H}
\end{equation}
where
\begin{equation}
\begin{split}
\hat{\mat{M}}^{\rm H} &=  \frac{ \hat{\mat{M}} + \hat{\mat{M}}^{\ast} }{2}\\
&=  \mat{C}^{-1}  \frac{ \mat{M}_{\chi}^{-1} - (\mat{M}_{\chi}^{-1})^{\ast} }{2} \succeq 0
\end{split}
\end{equation}
for any passive material ($\Im{}\chi \leq 0$).

The positivity of the scattered \eqref{P_sca2} and radiated \eqref{P_rad_final2} power is guaranteed if
\begin{equation}
\mat{B}^{\rm H} = \mat{A}_{\rm sca}^{\rm H} \preceq 0.
\end{equation}
The negative definiteness of $\op{A}_{\rm sca}$ carries over to $\mat{A}_{\rm sca}$, as explained in \cite{Rodriguez2013}.

Finally, in the case of the extinction power, positivity is straightforward by definition: the extinction power is the sum of the absorbed and scattered power, which we showed above that are positive.

\section{Computational Validation}
\begin{figure}[t]
\begin{center}
\includegraphics[scale=0.5]{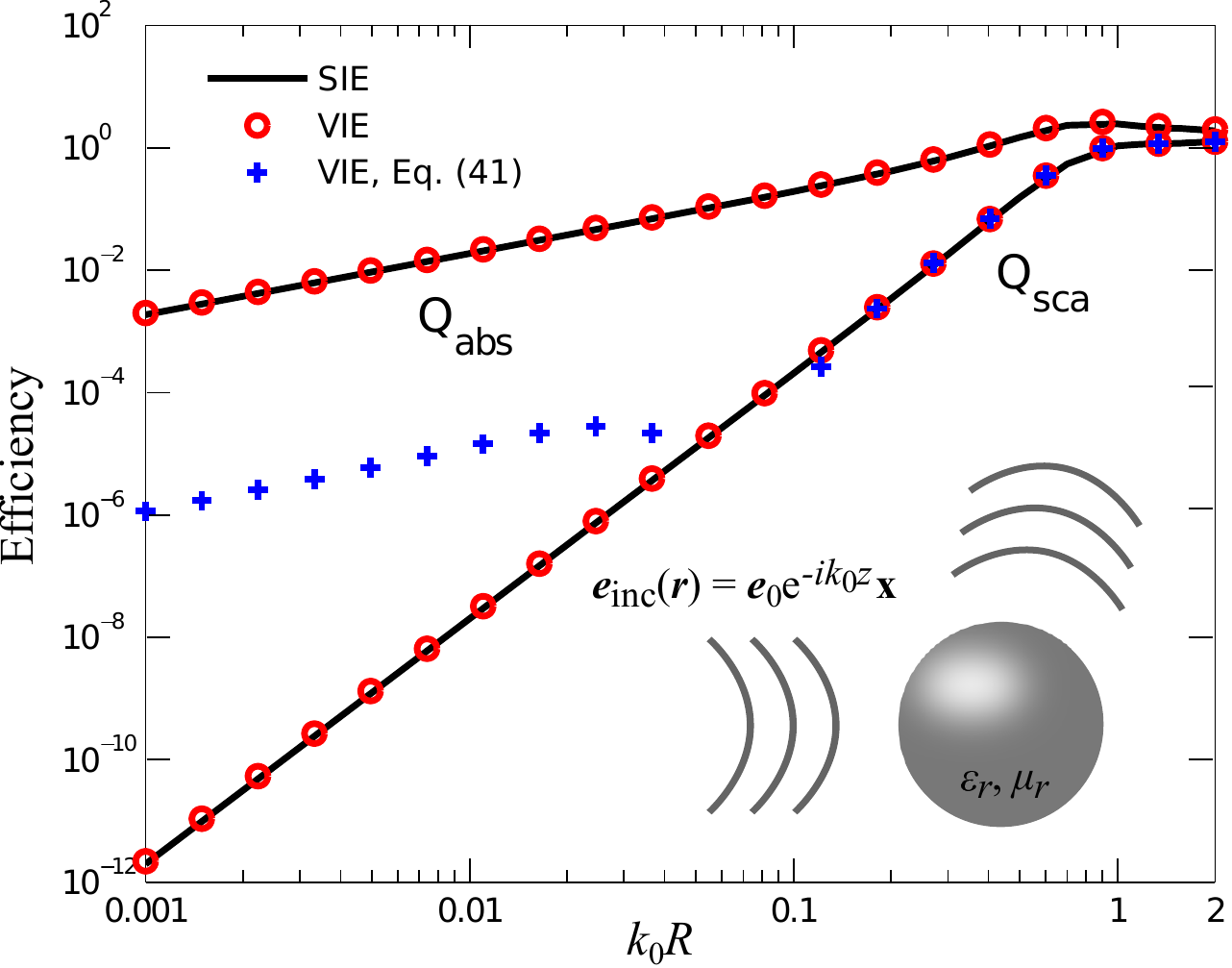}
\caption{Efficiencies for the case of a spherical particle of radius $R=1\, \rm \mu m$ irradiated by a plane wave. The particle is composed of material with $\epsilon_r = 3-6i$, $\mu_r = 2-1i$. The number of voxels used for VIE is $N=20^3$, resulting in $6N$ unknowns.}\label{fig_pw_VIEvsSIE}
\end{center}
\end{figure}

In what follows, we validate the new formulas by using them to compute some representative test cases. The JM-VIE formulation (referred herein as VIE, for simplicity) is numerically solved by means of an in-house FFT-based fast solver \cite{Polimeridis2013b}. More specifically, the unknown equivalent polarization and magnetization currents are approximated by a series of piecewise constant basis functions for each Cartesian component, with the support of each member of the discrete set being a voxel. A uniform grid of $N$ voxels is used for the discretization of the box that encloses the objects under study\footnote{Of course, one could choose different schemes for the numerical solution of the VIE method, e.g. based on a tetrahedral mesh coupled with a FMM solver \cite{Jarvenpaa2013}.}. The arising 6-D singular Galerkin inner products in \eqref{KN_op} are first reduced to 4-D singular (of lower order) integrals over the surfaces of the voxels \cite{Polimeridis2013b}, and then computed by means of \texttt{DIRECTFN} open-source package \cite{Polimeridis2013c,DIRECTFN}. The benchmark results are obtained with analytical formulas (Mie theory), and with a surface integral equation (SIE) method, and more specifically with the open-source package \texttt{scuff-em} \cite{scuff1, scuff2}. The default choice for the results regarding the VIE method are the boxed formulas presented above.

\begin{figure}[t]
\begin{center}
\includegraphics[scale=0.5]{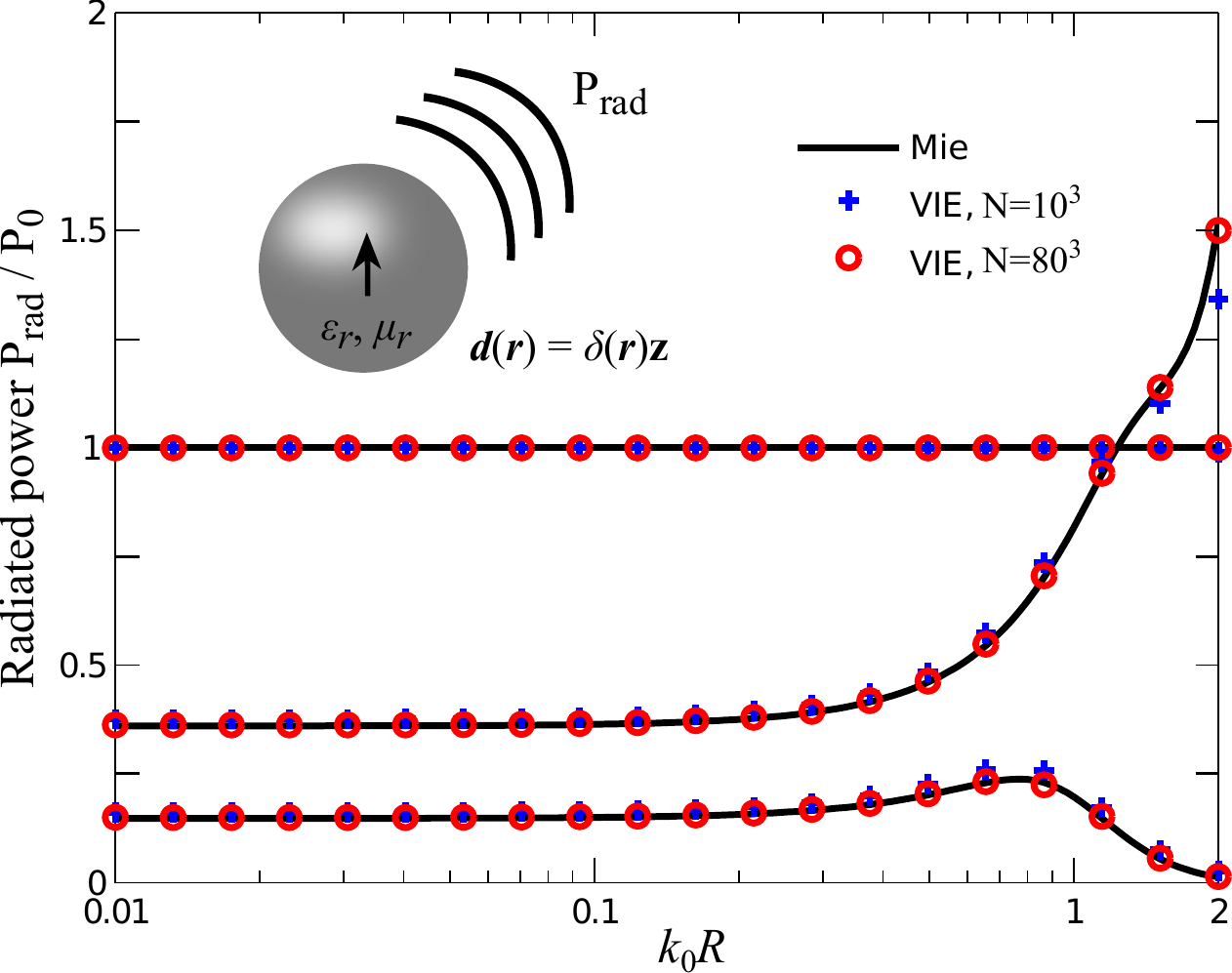}
\caption{Radiated power, normalized with respect to the power radiated in free-space, for the case of a spherical non-magnetic ($\mu_r=1$) particle of radius $R=1\, \rm \mu m$ irradiated by a Hertzian electric dipole with $z$-polarization, located at the center of the sphere. The electric permittivity is equal to $\epsilon_r=3-6i$, $\epsilon_r=3$, and $\epsilon_r=1$, as we go from the bottom to the top lines, respectively.}\label{fig_2}
\end{center}
\end{figure}

\subsection{Plane Wave Excitation}

We begin by computing efficiencies for scattering and absorption of a spherical particle of radius $R=1$ $\rm \mu m$ irradiated by a linearly-polarized $z$-traveling plane wave with electric field:
\begin{equation*}
\Ei = e_0  e^{-ik_0z} \univec{x}.
\end{equation*}
Fig. \ref{fig_pw_VIEvsSIE} plots efficiencies $Q^{\rm abs, sca}$ as functions of the dimensionless ``size parameter'' $k_0 R$. Efficiencies are obtained from cross sections $\sigma^{\rm abs, sca} = P^{\rm abs, sca}/P^{\rm inc}$, where $P^{\rm inc} = \frac{|e_0|^2}{2Z_0}$, by dividing by the geometrical cross section ($\pi R^2$) of the sphere, $Q = \sigma/(\pi R^2)$. The results are in good agreement with the efficiencies obtained by integrating the associated Poynting vectors by means of a SIE method \cite{Reid2014}. For the computation of the scattering efficiency, we present additional results based on the difference formula \eqref{P_sca1}, where one can identify the expected instabilities (blue +)  discussed in Section IIIC. Note that the two missing data points assume negative values; there is no guarantee for the positivity of the difference formula, as is the case in the PM.

\subsection{Dipole Excitation}

\begin{figure}[t]
\begin{center}
\includegraphics[scale=0.5]{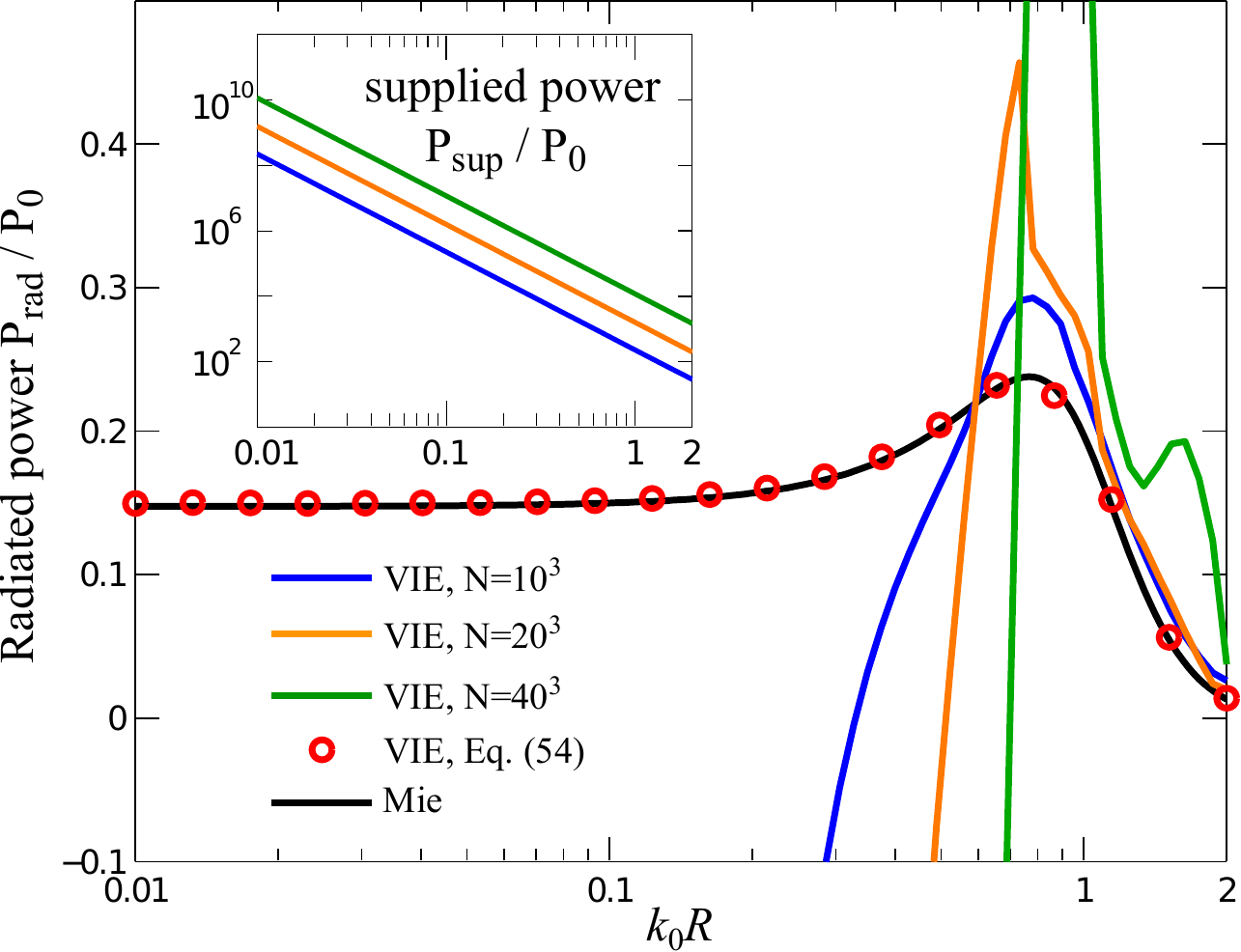}
\caption{Radiated power for the lossy case of Fig. \ref{fig_2} ($\epsilon_r = 3-6i$, $\mu_r=1$ ) using the difference formula \eqref{P_rad_final} (solid lines, except the black one). The supplied power $P_{\rm sup}$ is depicted in the inset.}\label{fig_3}
\end{center}
\end{figure}

We now consider the same sphere but irradiated by a Hertzian dipole (with moment equal to 1) directed along $z$-axis, and located at the center of the sphere. Fig. \ref{fig_2} plots the radiated power flowing through the surface of the spherical particle, normalized with respect to the power radiated by the dipole in free-space ($P_o$). We validate our VIE results by comparing against a reference analytical solution obtained via a Mie series\cite{Jackson_book},
\begin{equation}
P_{\rm Mie}=\frac{Z_0 c^2 a^{12}}{2 k_0^2\sqrt{6\pi}}\frac{| e^{ia} n^3|^2}{|D|^2}
\end{equation}
where
\begin{equation*}
\begin{split}
D &= \left[ n^2(a^3+a-i)-a +i \right] \sin{(na)}\\
&+ na \left[ -in^2(-1+a(a-i)) + a-i\right]\cos{(na)}
\end{split}
\end{equation*}
with $c$ being the speed of light, $n=\sqrt{\epsilon_r}$, and $a=k_0R$. Note that the power radiated by the dipole in free space $P_0=P_{\rm Mie}|_{n=1}$.

We consider three different scenarios in Fig. \ref{fig_2}, corresponding to different values of $\epsilon_r$ (with $\mu_r=1$). Specifically, we consider spheres with  $\epsilon_r=1$, $\epsilon_r=3$, and $\epsilon_r=3-6i$, corresponding to free-space, lossless, and dissipative media, respectively. As evidenced by the results, the quadratic formula \eqref{P_rad_final2} is stable both for low and high frequencies, where convergence to the exact solution is attained as the mesh discretization gets finer. Note that due to the uniform mesh used in this work, low resolution meshes suffer from staircase approximation errors.

Next, we consider the same dipole radiation but using the difference formula of \eqref{P_rad_final}, and show that it leads to the aforementioned instabilities (Section IIID). Fig. \ref{fig_3} plots the radiated power $P_{\rm rad} = P_{\rm sup}-P_{\rm abs}$ as obtained from \eqref{P_rad_final} (solid lines), for the case of the $\epsilon_r=3-6i$ lossy dielectric sphere of Fig. \ref{fig_2}. Also shown are the corresponding results from the Mie (black line) and VIE \eqref{P_rad_final2} (open circles) solutions. As discussed in Section IIID and \cite{Tai2000}, both the supplied and the absorbed power are infinite in this case. Nevertheless, the radiated power flowing through the surface of the sphere is the finite quantity presented in Fig. \ref{fig_2}. As depicted in the inset of Fig. \ref{fig_3}, the supplied and absorbed (not shown) powers diverge with the resolution of the mesh as $P \sim 1/\Delta V$. Consequently, the difference $P_{\rm sup}-P_{\rm abs}$ suffers from catastrophic cancellations that render the difference formula \eqref{P_rad_final} practically useless. This result highlights the importance of the quadratic formula \eqref{P_rad_final2}, which is remarkably stable and identically positive.

\begin{figure}[t]
\begin{center}
\includegraphics[scale=0.5]{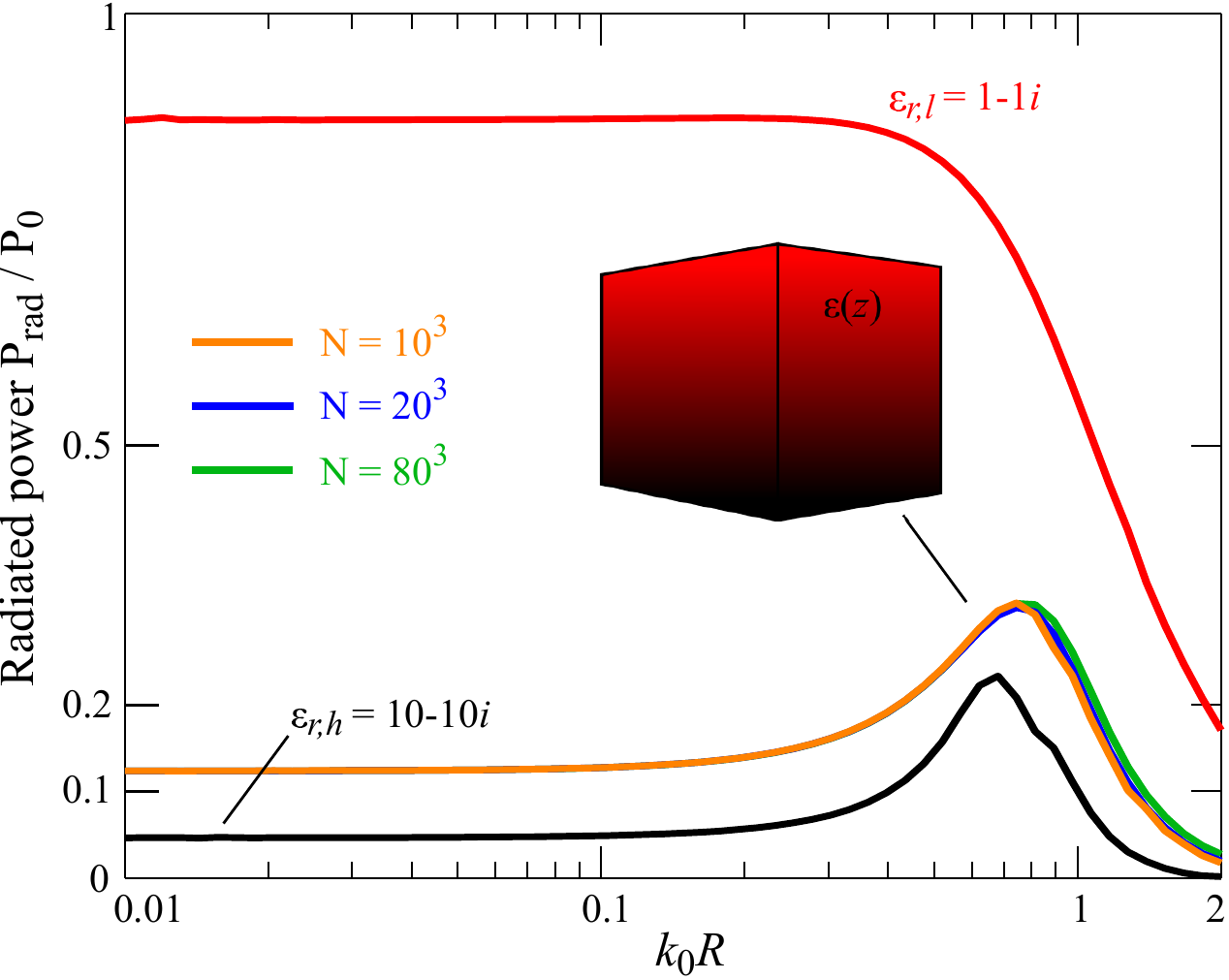}
\caption{Radiated power for the case of an inhomogeneous dielectric cube of length $L=2R$, irradiated by a $z$-directed dipole placed at the center. The relative permittivity varies linearly along the $z$-axis, i.e. $\epsilon_r \equiv\epsilon_r(z)$, as in \eqref{er_inhCUBE}. In addition, results for homogeneous cubes with the lowest and highest values of $\epsilon_r$ are presented.}\label{fig_4}
\end{center}
\end{figure}

Finally, we consider the case of an inhomogeneous dielectric cube irradiated by a $z$-directed dipole placed at the center. The continuous profile of the permittivity is given as follows:
\begin{equation}\label{er_inhCUBE}
\epsilon_r(z) = \epsilon_{r,l} + \frac{z+R}{2R}(\epsilon_{r,h}-\epsilon_{r,l}),\quad z \in [-R,R]
\end{equation}
where $\epsilon_{r,l}=1-1i$ and $\epsilon_{r,h}=10-10i$. The radiated power presented in Fig. \ref{fig_4} is computed using the quadratic formula \eqref{P_rad_final2} and converges as we refine the discretization. This is a particularly interesting example, since the continuous profile of the inhomogeneity rules out methods based on SIE formulations.

\section{Conclusion}

A collection of simple and stable formulas is presented for the computation of absorbed, scattered, extinction, and radiated power in VIE formulations. The proposed formulas (boxed equations in the manuscript) are accurate in a wide range of frequencies, and are based solely on volumetric quantities found in the associated linear system of equations. In addition, they preserve the positivity of the computed power, thus accurately capturing the physics of the problem. Thus, there is no need for significant post-processing, such as the evaluation of the fields and the integration of the Poynting vector along enclosing surfaces. By construction, the presented scheme is immune to the well-known instability issues that occur in Poynting's method. The efficient and compact absorption/radiation formulas presented herein are expected to be especially useful---besides applications in classical scattering/radiation problems---in computations of EM fluctuation phenomena, including radiative heat transfer and Casimir forces between complex bodies. Our analysis is based on a current-based VIE formulation, but similar formulas may be easily derived, with only minor modifications, for the case of VIE formulations based on fields or fluxes.

\section{Acknowledgments}

This work was supported in part by grants from the Singapore-MIT programs in Computational Engineering and in Computational and Systems Biology, from the Skolkovo-MIT initiative in Computational Mathematics, and from the Army Research Office through the Institute for Soldier Nanotechnologies under Contract No. W911NF-07-D0004.

\appendix[On the analyticity of $P_{\rm ext}$]

\section{On the analyticity of $P_{\rm ext}$}

As mentioned in Section IIIB, it is important to write the formula for the extinction power \eqref{P_ext} in a form where the analyticity (in the lower half of the complex-$\omega$ plane) is shown explicitly, so as to be able to exploit it both from a theoretical and a practical perspective. Indeed, \eqref{P_ext} can be written with the help of \eqref{rhs} and \eqref{hatM} as follows:
\begin{equation*}
\begin{split}
P_{\rm ext} &= \frac{1}{2}\Re{} \begin{pmatrix} \matvec{e}_{\rm inc} \\ \matvec{h}_{\rm inc}  \end{pmatrix}^{\ast} \matvec{x} \\
&= \frac{1}{2}\Re{} \begin{pmatrix} \matvec{e}_{\rm inc} \\ \matvec{h}_{\rm inc}  \end{pmatrix}^{\ast} \mat{W} \begin{pmatrix} \matvec{e}_{\rm inc} \\ \matvec{h}_{\rm inc}  \end{pmatrix}
\end{split}
\end{equation*}
where $\mat{W}=\mat{A}^{-1}(\mat{C}\, \mat{M}_{\chi})$ is the matrix arising from the discretization of the operator $\opd{W}$ relating the incident fields to the induced currents, i.e.,
 \begin{equation*}
\begin{pmatrix}
\J     \\
\M
\end{pmatrix}
=
\opd{W}
\begin{pmatrix}
\Ei     \\
\Hi
\end{pmatrix}.
\end{equation*}
Causality implies that $\opd{W}$ is an analytic function in the lower half of the complex-$\omega$ plane \cite{Landau_book}.  Alternatively, it is sufficient to assume passivity rather than causality, since the former implies the latter in a time-invariant linear system \cite{Welters2014}.

Finally, we can eliminate the complex conjugation by exploiting the conjugate symmetry of the Fourier transform of any real incident field:
\begin{equation*}
\begin{split}
P_{\rm ext} = \frac{1}{2}\Re{} \begin{pmatrix} \matvec{e}_{\rm inc}(-\omega) \\ \matvec{h}_{\rm inc}(-\omega)  \end{pmatrix}^T \mat{W}(\omega) \begin{pmatrix} \matvec{e}_{\rm inc}(\omega) \\ \matvec{h}_{\rm inc}(\omega)  \end{pmatrix}.
\end{split}
\end{equation*}
As desired, this is the real part of an analytic function in the lower-half complex-$\omega$ plane, as long as the incident fields are entire (everywhere-analytic) functions of $\omega$ (which is true for all typical incident fields, such as planewaves, gaussian pulses, or any pulse that is compactly supported in the time domain \cite{Strichartz_book}).

\bibliographystyle{IEEEtran}
\bibliography{IEEEabrv,References}

\end{document}